\documentclass[preprint,12pt]{elsarticle}

\usepackage{lineno,hyperref,amsmath}

\journal{Nuclear Instrumentations and Methods in Physics Research A}

\newcommand{\nqq[1]}{10^{#1} } 
\newcommand{\nq[1]}{\cdot 10^{#1} \; n_{eq}/cm^2 } 
\begin{document}

\begin{frontmatter}

\title{Silicon Sensors for Future Particle Trackers}

\author[1]{N. Cartiglia\corref{corrauth}}
\cortext[corrauth]{Corresponding author}
\ead{cartiglia@to.infn.it}

\author[1,2]{R.~Arcidiacono}
\author[5,6]{G.~Borghi}
\author[5,6]{M.~Boscardin}
\author[1,4]{M.~Costa}
\author[3]{Z.~Galloway}
\author[1,4,7]{F.~Fausti}
\author[1,2]{M.~Ferrero}
\author[5,6]{F.~Ficorella}
\author[1]{M.~Mandurrino}
\author[3]{S.~Mazza}
\author[1,4]{E. J.~Olave}
\author[5,6]{G.~Paternoster}
\author[1,4]{F.~Siviero}
\author[3]{H. F-W.~Sadrozinski}
\author[1,4]{V.~Sola}
\author[1]{A.~Staiano}
\author[3]{A.~Seiden}
\author[1,4]{M.~Tornago}
\author[3]{Y.~ Zhao}

\address[1]{INFN, Torino, Italy}
\address[2]{Universit\`a del Piemonte Orientale, Italy}
\address[3]{SCIPP, University of California Santa Cruz, CA, USA}
\address[4]{Universit\`a di Torino, Torino, Italy}
\address[5]{Fondazione Bruno Kessler, Trento, Italy}
\address[6]{TIFPA-INFN, via Sommarive 18, 38123, Povo (TN), Italy}
\address[7]{now at DE.TEC.TOR. Devices \& Technologies Torino S.r.l., Torino, Italy}

\begin{abstract}
Several future high-energy physics facilities are currently being planned. The  proposed projects  include high energy $e^+ e^-$ circular and linear colliders, hadron colliders and muon colliders, while the Electron-Ion Collider (EIC) has already been approved for construction at the Brookhaven National Laboratory.  Each proposal  has its own advantages and disadvantages in term of readiness, cost, schedule and physics reach, and each proposal requires the design and production of specific new detectors. This paper first presents the performances required to the future silicon tracking systems at the various new facilities, and then it illustrates a few  possibilities for the realization of such silicon trackers. The challenges posed by the future facilities require a new family of silicon detectors, where  features such as impact ionization, radiation damage saturation, charge sharing, and analog readout are exploited to meet these new demands. 
\end{abstract}

\begin{keyword}
  Silicon \sep Fast detector \sep Low gain \sep Charge multiplication \sep LGAD
  \MSC[2018] XX-XX\sep  XX-XX
\end{keyword}

\end{frontmatter}



\section{Introduction} 
Several future facilities of high-energy physics are presently been considered, with a timescale of 15 - 30 years. The proposals consider accelerators for $e^+ e^-$, hadrons, muons and electron-ions.  The requirements for silicon trackers differ mostly upon the type of particles that are accelerated, and not on the specifics of a given proposal. For examples, at the various $e^+ e^-$ machines  (Circular Electron Positron Collider - CPEC, Compact Linear Collider - CLIC, Future Circular Collider - FCC-ee, and International Linear Collider  - ILC) the key requests are  about low material budget and very good spatial resolution, with limited requests for radiation resistance or precise timing ($\sigma_t < 50\;ps$). On the other hand, at hadron machines (Future Circular Collider - FCC-hh, High Energy LHC - HE-LHC, and Super Proton Proton Collider - SppC) the most challenging requests are the radiation resistance  (fluences above $1\nq[17]$) and the spatial and time precision (pileup $\sim$ 1000 events/bunch crossing, $\sigma_t \sim 5 \; ps/hit,  \sigma_x \sim 5 \; \mu m/hit$). The requests for the muon collider are similar to those of CLIC, however, plus a time resolution of $\sim$ 50 ps for the inner tracker and $\sim$ 100 ps for the outer tracker. 
 Table~\ref{table:silreq}, taken from~\cite{Aleksa:2649646}, summarizes the present requirements for the silicon trackers at various facilities,  while an updated review has been presented at the TREDI 2020 conference ~\cite{TREDI2020}.  There are several possible paths to future silicon trackers~\cite{Aleksa:2649646}\cite{HSTD12M}, including HVCMOS, low field monolithic sensors, and hybrid detectors. In the following part of this paper, three key aspects of future silicon trackers will be considered: (i) extension of  picosecond time resolution to fluences above the present limit of 1-2$\nq[15]$, (ii) design of silicon sensors able to withstand fluences in the range 1-10$\nq[16]$, and (iii) capability of obtaining very good position resolution without increasing dramatically the channel count.  The technological challenges presented are connected to the design of the silicon sensors, however, it is important to stress the importance of the interconnection with the front-end electronics:   silicon sensors and associated electronics succeed or fail together.  In ~\cite{HSTD12lai}, the evolution of 3D sensors to meet the requirement of 4D tracking is presented: equivalently  to the present situation of silicon trackers, 3D sensors will be very important to cover the area with the most extreme fluence levels. This topic is not further developed in the present contribution. 

\begin{table}[h]
\begin{center}
\begin{tabular}{|l|c|c|c|c|c|c|}
\hline
&	 HL-LHC &	SPS & FCC-hh &  FCC-ee & CLIC & mu Col.  \\ \hline \hline
Fluence & $\nqq[16]$& $\nqq[17]$& $\nqq[17]$& $<\nqq[10]$& $<\nqq[11]$ & \\
$[n_{eq}/cm^2/y]$  & & &  & & &  \\
Hit rate $[s^{-1}cm^{-2}]$  & 2-4G & 8G&   20 G& 20 M & 240 k&  \\
Inn. tracker $[m^{2}]$ & 10 & 0.2 &   15 G& 1 M & 1 &  \\
Out. tracker $[m^{2}]$  & 200 & - &   400 G& 200 & 140 &  \\
Pixel size $[\mu m^2]$  & 50x50 & 50x50 &   25x50 & 25x25 & 25x25 &  \\
Time res $[ps]$ & 50  & 40 &   10& 1k & 5k & 50-100  \\ \hline 
\end{tabular}
\caption{Summary of the parameters of future silicon trackers at new facilities}
\label{table:silreq}
\end{center}
\end{table}

\section{Extension of UFSD picosecond time resolution to fluences above the present limit of 1-2$\nq[15]$ }
\label{sec:sig}
\begin{figure}[htb]
\begin{center}
\includegraphics[width=0.95\textwidth]{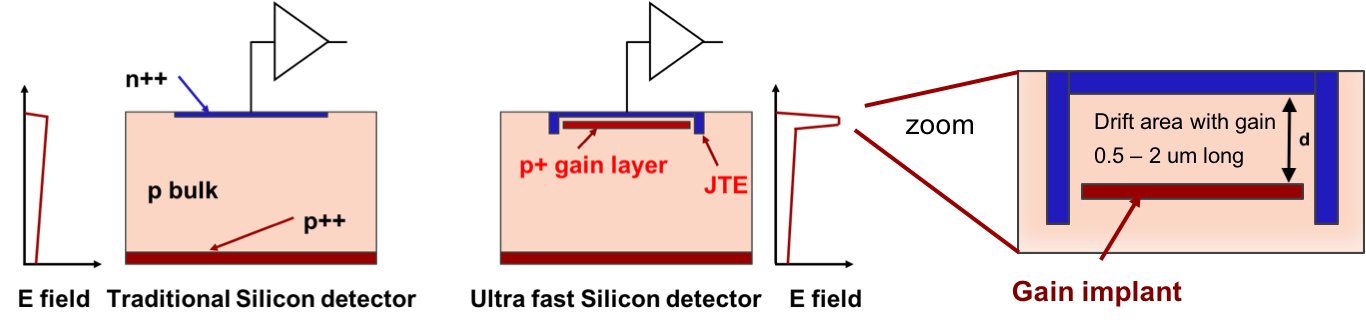}
\caption{Key layout features of an n-in-p silicon sensor (left side) and of an UFSD (center). The right side shows an expanded view of the multiplication region.}
\label{fig:ufsd}
\end{center}
\end{figure}

In the last 5 years, silicon detectors have gone from being considered unfit to perform accurate timing measurement (with precision $\sigma_t < 50 \; ps$) to being the only viable solution for the  construction of large tracker detectors performing the concurrent measurements of space and time, the so called 4D-tracking system~\cite{ROPP}. This change of paradigm was brought about by the introduction of  low gain avalanche diodes (LGAD) ~\cite{LGAD1} and their subsequent design optimization for timing application (Ultra Fast Silicon Detector, UFSD)~\cite{UFSD1}.  Figure~\ref{fig:ufsd} illustrates the key technological steps of this evolution: to the design of a traditional n-in-p sensor, left side of the picture, an additional deep p-implant has been added (central part of the picture), so that in the region between this implant and the $n^{++}$ read-out electrode, the electric field is high enough (right side of the picture) to generate multiplication of the drifting electrons. Presently, LGAD are manufactured by several foundries, including CNM (Spain)~\cite{Carulla:2019bsa}, FBK (Italy)~\cite{HSTD12Arci}, Hamamatsu ~\cite{HSTD12HPK} (Japan), Micron (England), BNL~\cite{Giacomini:2018tbt} (USA), and NDL~\cite{HSTD12NDL} (China).  

\begin{figure}[htb]
\begin{center}
\includegraphics[width=0.95\textwidth]{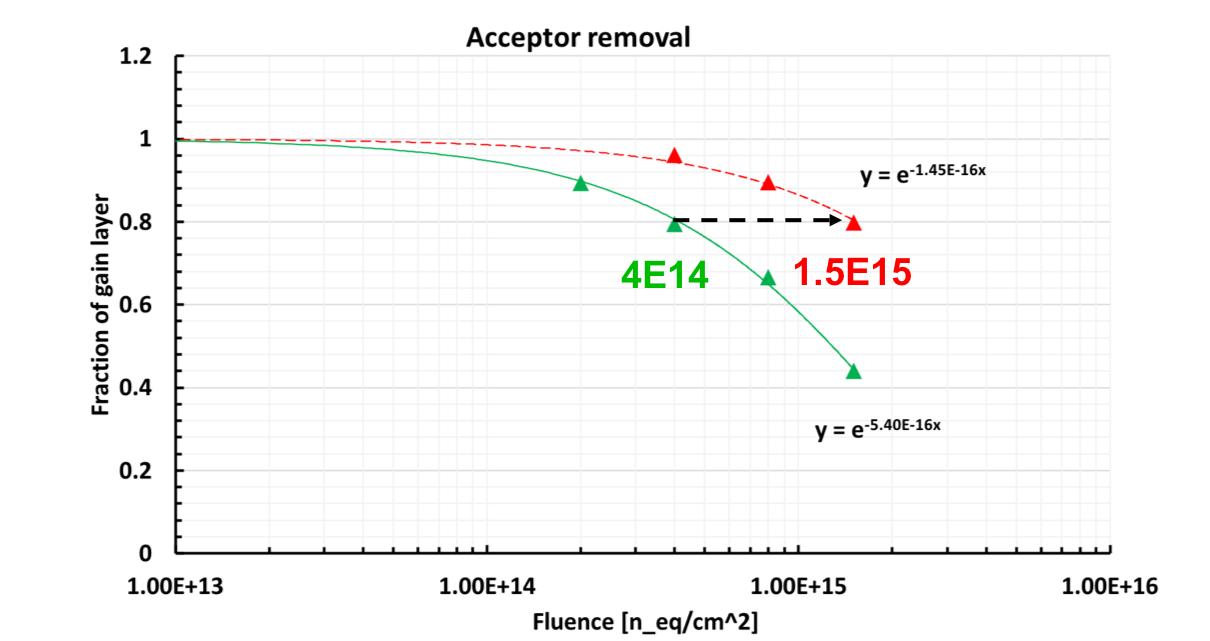}
\caption{Active fraction of gain layer in UFSD as a function of irradiation. The green curve represents the typical behavior for prototypes manufactured in 2016 while the red curve for those manufactured by FBK in 2018 with carbon infusion.}
\label{fig:ar}
\end{center}
\end{figure}

The defining feature of the UFSD design, the deep $p^{++}$ implant, responsible to generate the high E-field needed to create controlled multiplication, has, at the moment, a radiation resistance limited to fluences of about 1-2$\nq[15]$. The underlying reason for this effect is the acceptor removal mechanism~\cite{HSTD12Sadro} ~\cite{1748-0221-10-07-P07006} that decreases the doping density of the gain layer to a level where it does not any longer generate a high-enough field. In the past 3 years, there has been a lot of development in the understanding of the acceptor removal mechanism and in the design of more radiation resistance UFSD. Figure~\ref{fig:ar} reports this progress by showing the active fraction of the gain layer as a function of fluence for two typical  FBK UFSD productions, one from 2016 and one from 2019. The  key technological difference between the two productions is the infusion of carbon in the gain layer, which reduces the acceptor removal mechanism ~\cite{FERRERO201916}\cite{Vertex19Moll}.
If we consider as limit of the radiation resistance the value at which the gain layer decreases by 20\%, between the 2016 and 2019 productions there is an improvement of a factor of  4, from 4$\nq[14]$ to 1.5$\nq[15]$. This is mostly due to the carbon infusion and to a better design of the gain layer. The 20\% value is based on experimental measurements ~\cite{Galloway}\cite{GK15}\cite{Mazarticle} demonstrating that, for reductions below this value, the electric field in the gain layer is too low to be restored by an increase of the detector bias.  

There are several UFSD productions planned for the next few years,  in conjunction with the ATLAS~\cite{Collaboration:2623663} and CMS~\cite{CMS:2667167} timing layer and for R\&D studies (mostly in connection with the CERN RD50 activities~\cite{RD50}).  There are presently two venues that are being explored to improve the UFSD radiation resistance: (i) decreasing the acceptor removal rate and (ii) enhancing the recovery capability of the bias voltage.  The first point,  increasing  the gain layer radiation resistance, is actively been pursed by FBK with the production of UFSD wafers using a varying density of infused carbon. In the  2019 FBK UFSD3 production, the density of carbon infusion used in the 2018 UFSD2 production has been increased by a factor of 2, 3 and 5, without finding any  improvement in radiation resistance~\cite{HSTD12Arci}. In the 2020 UFSD3.2 production, the density of carbon infusion has been reduced to 80\% and 40\% of that of UFSD2. This production  will therefore complete the scan in carbon density and will help pinpoint the dose of carbon infusion that maximizes the radiation resistance. Acceptor removal can possibly be decreased by the addition of different elements, besides carbon: the RD50 collaboration is pursuing this path by investigating the microscopic mechanism of acceptor removal and modeling the beneficial effects of carbon.   

\begin{figure}[htb]
\begin{center}
\includegraphics[width=0.7\textwidth]{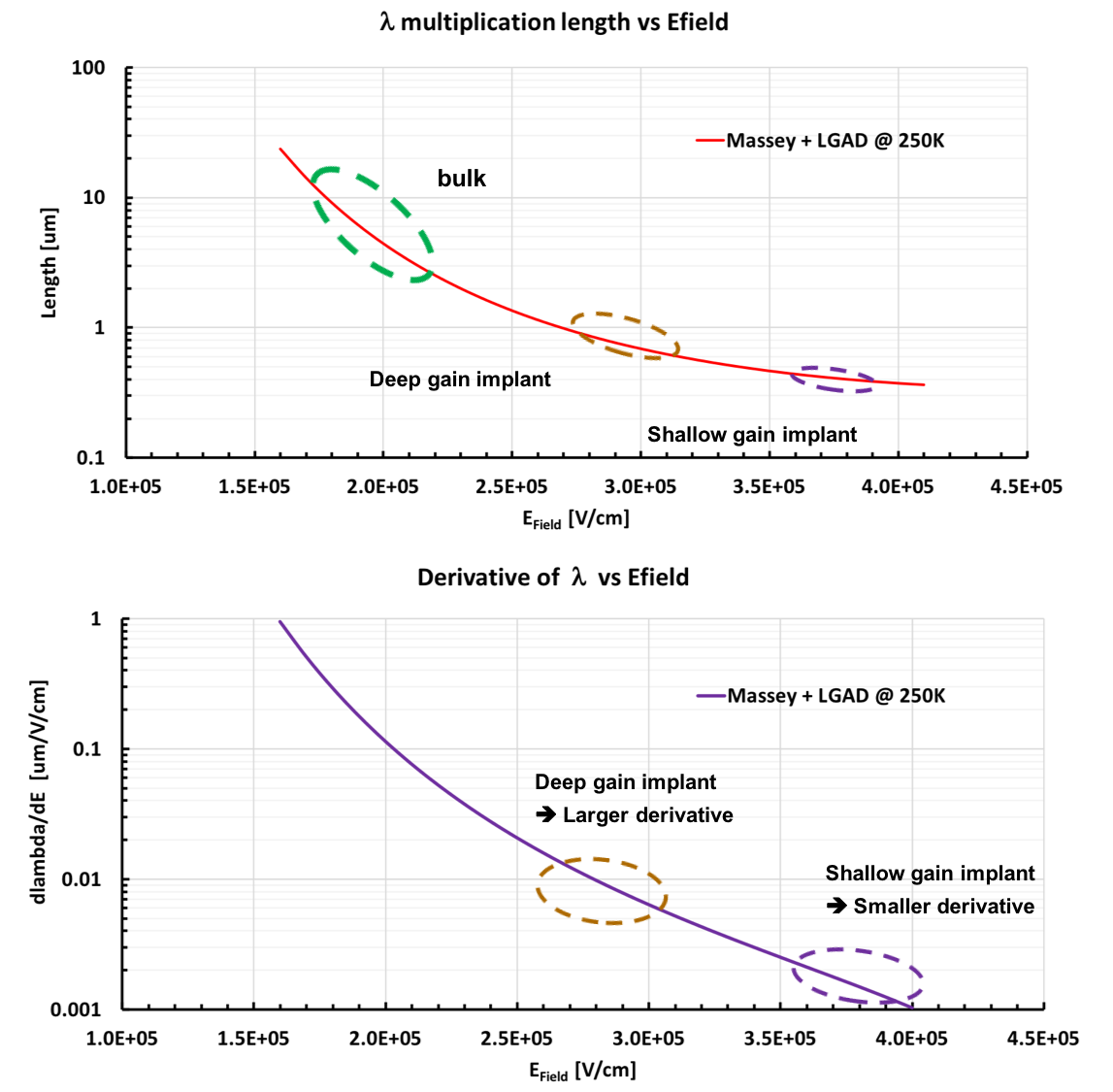}
\caption{Top: multiplication length $\lambda$ as a function of the electric field. Bottom: derivative of the multiplication length as a function of the electric field. }
\label{fig:field}
\end{center}
\end{figure}

The second technique to increase the radiation resistance of the UFSD design is to enhance the recovery capability of the bias voltage: in UFSD, as the gain layer is deactivated by radiation, the electric field in the gain region is kept high by increasing the bias voltage. The field per micron is linear with the bias voltage and inversely proportional to the sensor thickness, $E = Bias/Thickness$.  One obvious choice is to make the sensor thinner:  a bias increase of 100 Volt in a $25\; \mu m$ thick sensor increases the field by 4V/$\mu m$ while only by 1V/$\mu m$ in a  $100\; \mu m$ thick sensor. The obvious drawbacks of this choice are that thin sensors have higher capacitance and generate smaller signal.  Another option is to design the gain layer such that  the bias voltage increase has a stronger impact on charge multiplication~\cite{NC-RD50-H}. Charge multiplication happens in the space between the gain layer and the $n^{++}$ read-out electrode, right panel of Figure~\ref{fig:ufsd}. The gain G is defined as 

\begin{equation}
\label{eq:gain}
G \propto e^{\alpha(E,T) \cdot d} \; \; with \;\; \alpha(E,T) \propto e^{-(a+b\cdot T)/E}
\end{equation}

where $d$ is the total distance and $\alpha(E,T)$ the impact ionization coefficient,  function of the field E and the temperature T via the two experimental parameters $a,b$. $\lambda = 1/\alpha$, represents the length to achieve $G  = e$.  The ratio $d/\lambda$ determines the gain: if  two gain layers are implanted at different depths, $d_1, d_2$, they will achieve the same gain when $d_1/\lambda_1 = d_2/\lambda_2$.  Figure~\ref{fig:field}, top panel,  shows the dependence of $\lambda$ upon the field, according to the Massey impact ionization model~\cite{Massey}: in deeper gain layer designs, the drift length $d$ is longer and  the electric field lower than for shallower gain layer.
The restoration power of the the bias voltage is evaluated by studying the derivative $d\lambda/dE$, bottom panel of  Figure~\ref{fig:field}: for very high fields, i.e. shallow gain layer, the derivative is very small, indicating that a large increase in bias is necessary to restore the field needed for multiplication while for deeper gain layer a much smaller increase is necessary. For sensors with deep gain layer, the recovering effect of the bias voltage is therefore much higher than for those with shallow gain layer.  This effect has been confirmed studying the recovering power of the bias voltage with increasing temperature (the gain goes down as the temperature goes up): sensors with deep gain layers require a  voltage increase of 1V/$^o$C to keep the gain constant  while sensors with shallow gain layer require almost an  increase of 2V/$^o$C (the coefficient $b$ is therefore different in the two cases).  A drawback of deep gain layer is that they need to be doped less, and therefore are more prone to the acceptor removal mechanism. In the 2020 FBK UFSD3.2 production, the combination of carbon infusion and deep gain layer will be explored. 

The different aspects presented in this section point to a potential extension of the radiation hardness of UFSD, hopefully above fluences of 5$\nq[15]$.

\section{Exploitation of radiation damage saturation  in the design of silicon sensors for  fluences above  1$\nq[16]$}

\begin{figure}[htb]
\begin{center}
\includegraphics[width=0.95\textwidth]{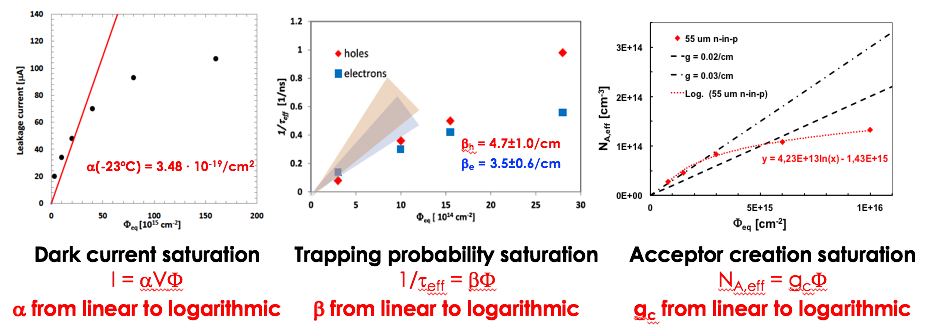}
\caption{Radiation damage in silicon sensors as a function of fluence: the leakage current, the trapping probability, and the creation of  acceptor-like states show clear signs of  saturation.}
\label{fig:sat}
\end{center}
\end{figure}

In the last few years,  a set of novel measurements on highly irradiated sensors (fluences $\sim 1\nq[17]$) have demonstrated that silicon sensors behave better after heavily irradiation than what was predicted by extrapolating lower fluence data ($\phi < 1\nq[15]$) to higher values~\cite{Kramberger:2019ydi}\cite{Cartiglia:2019eut}\cite{TREDI2020M}. 
Figure~\ref{fig:sat} (taken from~\cite{Vertex19Sola} and reference therein) exemplifies this saturation effect for 3 different parameters:  the leakage current, the trapping probability, and the creation of  acceptor-like states. 

As the three panels show, the initial linear dependence of the damage with fluence becomes a logarithmic trend at larger fluence. The reason for this change is not understood. A naive consideration is that after a fluence of 1$\nq[15]$ every single silicon lattice cell  has been traversed by a particle: for fluences above 1$\nq[15]$, radiation damage happens to already damaged cells and, possibly, damage on damaged silicon has less consequences.  The exploitation of saturation effects is the key to the design of silicon sensors able to work at fluences about 1$\nq[17]$: our current understanding is that, once the saturation effects are included, thin sensors will continue to work. Even after 1$\nq[17]$ the changes to thin silicon sensors (20-30 $\mu$m) are not dramatic: the leakage current is quite low, the charge collection efficiency is high and the full  depletion voltage,   
 $V_{FD} = e|N_{eff}|x^2/2\epsilon$, where $N_{eff}$ is the bulk doping and $x$ the detector thickness, remains below 500-600V.  The drawback of thin sensors is that the generated signal is too low: present state-of-the-art ASICs, for example those  produced for HL-LHC,  require a minimum charge of about 1 fC~\cite{TREDI2019_LD}. This problem could be solved by using sensors with internal gain, however, gain in very irradiated sensors has not been studied enough to know if this approach might or might not work. Impact ionization in thin sensors should happen in the bulk, at relative low fields, as indicated in Figure~\ref{fig:field}. In the current models of impact ionization available in TCAD\footnote{www.synopsys.com/silicon/tcad.html}, the impact ionization coefficient $\alpha$ does not have an explicit dependence upon the fluence $\phi$, however, it can be added by simply duplicating the dependence upon temperature:

\begin{equation}
\label{eq:que}
\alpha(E,T) \propto e^{-(a+b\cdot T)/E}\rightarrow  \alpha(E,T,\phi) \propto e^{-(a+b\cdot T+c\cdot \phi)/E}.
\end{equation}

\begin{figure}[htb]
\begin{center}
\includegraphics[width=0.7\textwidth]{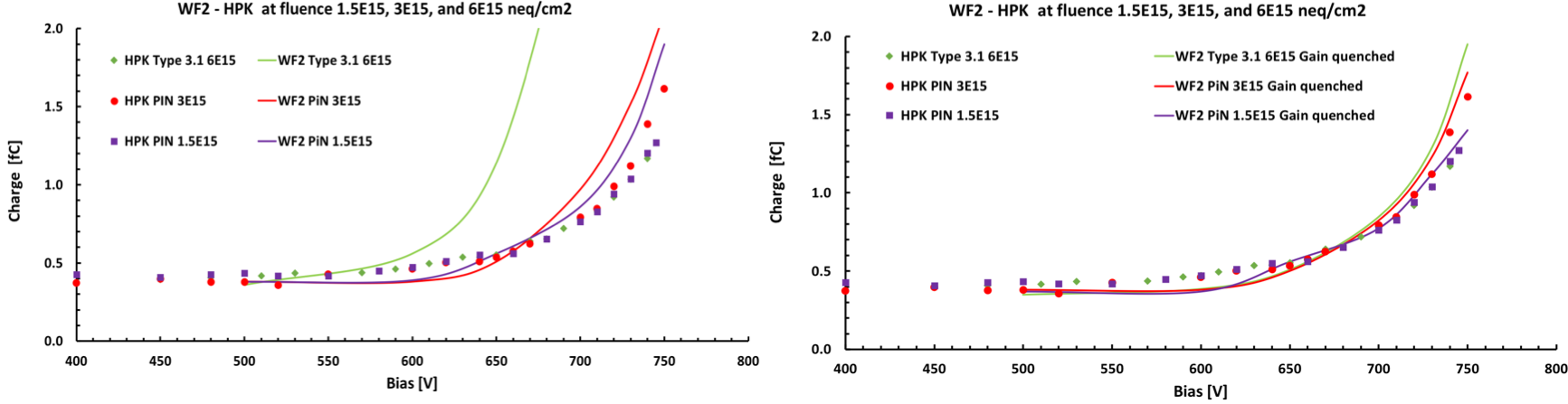}
\caption{Signal integral in 45-$\mu$m thick HPK sensor as a function of the bias voltage for 3 irradiation levels. The solid curve shows the predicted charge without (top) and with (bottom) gain quenching.}
\label{fig:bulk}

\end{center}
\end{figure}

As initial study, the multiplication in the sensor bulk has been investigated for HPK 45-$\mu$m thick sensors , irradiated  up to 6$\nq[15]$ and compared with the Massey impact ionization model as implement in the Weightfield2 (WF2) simulation program~\cite{WF2}. The left side of Figure~\ref{fig:bulk} shows the collected signal as a function of bias voltage for 3 fluences (1.5, 3, and 6$\nq[15]$), together with the prediction of WF2.  Note that the sensor irradiated at 6$\nq[15]$ is a UFSD and the left over gain from the gain layer is taken into account in the simulation. 
The gain simulation, in absence of gain quenching ($c = 0$), predicts an increasing gain with fluence, driven by the field generated by the bulk doping. This prediction is clearly not supported by the data, Figure~\ref{fig:bulk} top panel. The simulation can be reconciled with the data introducing a  quenching mechanism, as that proposed in equation~\ref{eq:que}, with $c = 2*10^{-11}V/\phi$. With this addition, the simulation and the data agree quite well, Figure~\ref{fig:bulk} bottom. This study demonstrates that gain is still present after fluence of 6$\nq[15]$, albeit, already quenched. The investigation of gain in thin sensors will continue in the next years by irradiating even thinner sensors (20-30 \;$\mu$m), where  bulk multiplication  might be less affected by lattice defects since it is achieved at higher field and lower $\lambda$.
Overall, radiation damage saturation suggests the possibility of using thin sensors for future FCC-hh trackers;  future studies of impact ionization in heavily irradiated sensors will shed light about the feasibility of this idea.

\section{Charge sharing as a solution for very good position resolution without using very small pixels} 
\begin{figure}[htb]
\begin{center}
\includegraphics[width=0.95\textwidth]{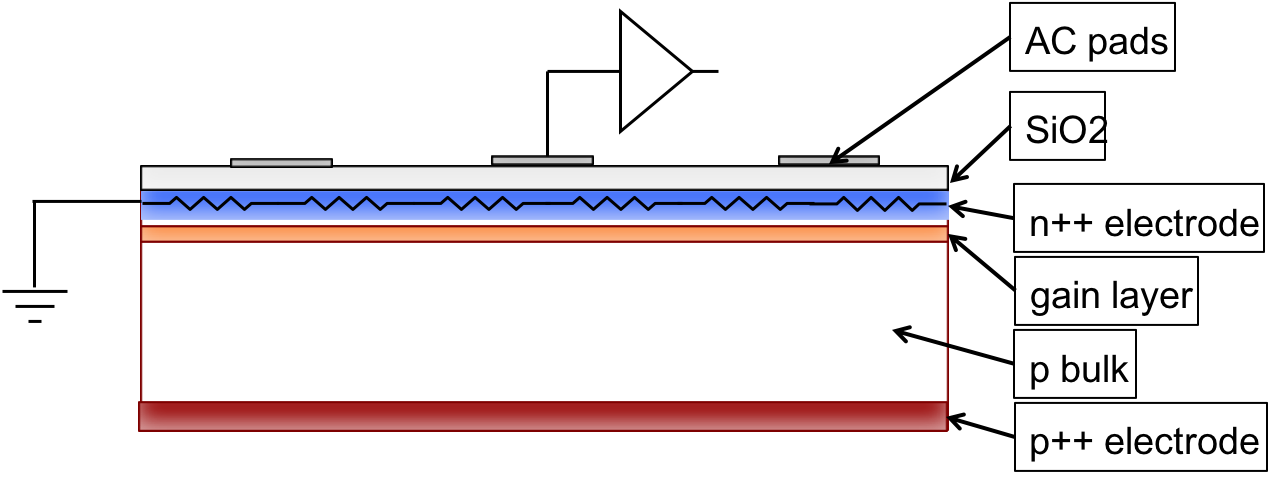}
\caption{Schematic of an a AC-LGAD sensor.}
\label{fig:rsd}
\end{center}
\end{figure}

Good position resolution is normally achieved by designing sensors with small pixel: in binary read-out, the resolution is normally quoted as $bin\;size/\sqrt{12}$.  The proposed future detectors, listed in Table~\ref{table:silreq}, have pixel sizes  from 50x50 to 25x25 $\mu m^2$; such a high granularity is clearly not optimal, unless the occupancy is also very high (as it is for example at FCC-hh).  Charge sharing between pads yields to a much more precise localization of the hit, however, the e/h drift lines in traditional pixel detectors are such that analog sharing is limited.  Charge sharing in silicon sensors can be obtained by designing a new type of device  where the signal on the read-out pads is not induced (following Ramo's theorem) during the drift of the e/h charge carriers in the bulk, but it is picked up in AC coupled mode during the propagation of the signal towards ground.  AC-coupled LGADs~\cite{TREDI2015N}\cite{HSTD11H} are designed on this principle, maximizing charge sharing between pads to obtain a position resolution a factor of 5-10 better than $bin\;size/\sqrt{12}$. AC-coupled LGAD, Figure~\ref{fig:rsd},  are n-in-p sensors, with a continuous gain layer, a resistive $n^{++}$ implant, and a thin dielectric layer for AC coupled read-out.  The size of the AC metal pads determine the readout segmentation and it can be adjusted to any geometry by simply changing two production masks (metal etching and overglass), leaving the rest of the sensor identical. The goal of the resistive $n^{++}$ layer is to keep the signal localized, to reduce the capacitance seen by the readout pad, and to induce the AC signal on the metal pad, somewhat equivalent to the role of the  graphite layer in the RPC.  For this reason, AC-LGAD are also called {\it resistive silicon detector} (RSD). AC-LGAD have been produced by CNM in 2017, by FBK within the RSD project\cite{8846722}\cite{Mandurrino:2020ukm}, and by BNL~\cite{Giacomini:2019kqz}.

\begin{figure}[htb]
\begin{center}
\includegraphics[width=0.95\textwidth]{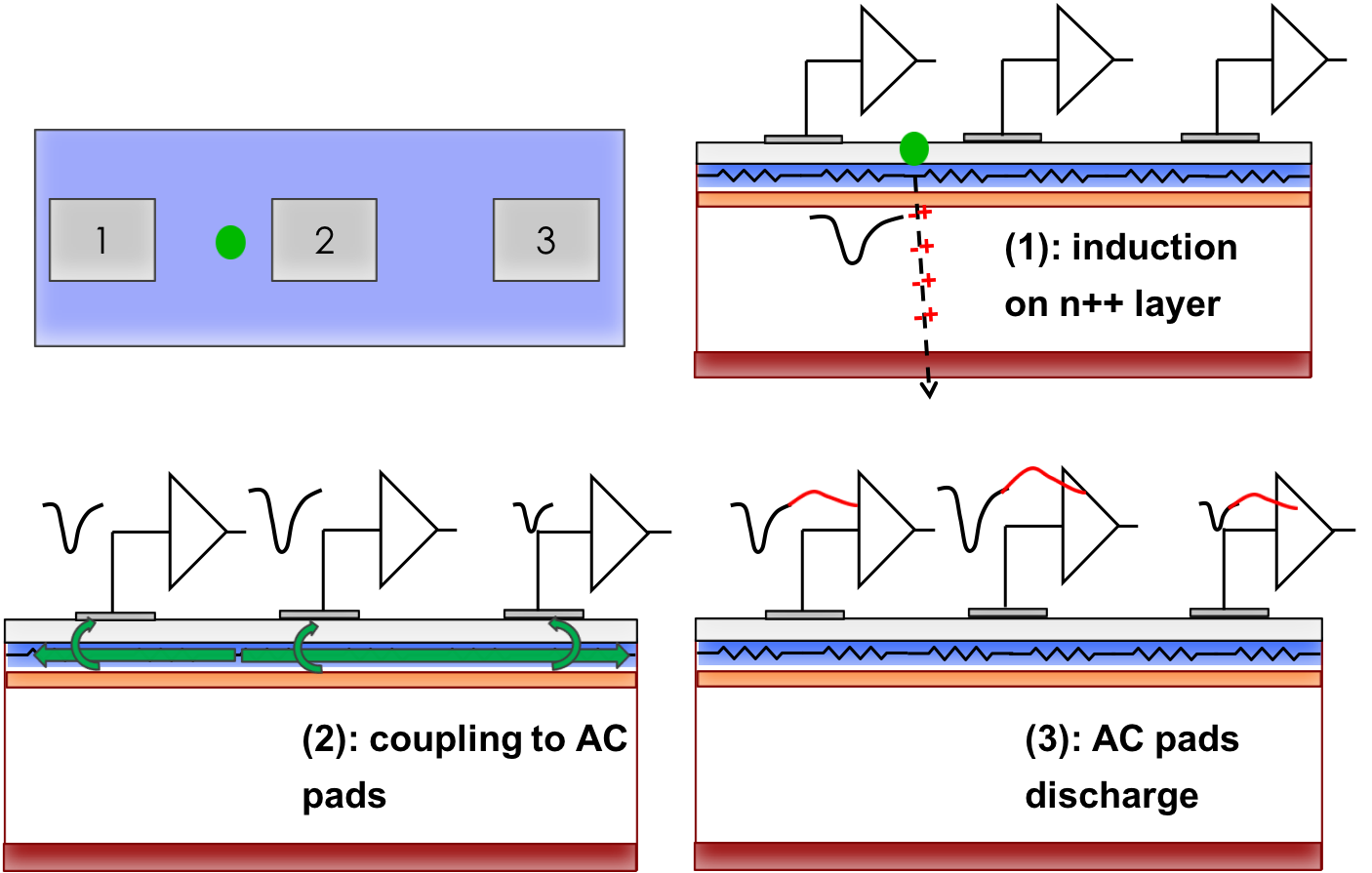}
\caption{Signal formation in AC-LGAD: the signal is seen on the electrodes with a delay proportional to the distance from the  impinging point indicating that the formation mechanism is not direct induction.}
\label{fig:al}
\end{center}
\end{figure}

Signal formation in AC-LGAD  happens in the 3 phases~\cite{TREDI2020N} sketched in Figure~\ref{fig:al}: (i) The first step is similar to all other silicon sensors: the drift of the e/h pairs generates an induced signal on the $n^{++} $electrode. Note that there is no direct induction on the metal pads, the $n^{++}$ is conductive enough to stop it.  (ii) The signal spreads laterally along the lossy transmission line composed by the $n^{++}$ layer and the bulk and AC capacitance. The metal pads act as pick-up electrodes and record a signal.  (iii) In the last phase, the AC pads discharge, with an RC that depends on the readout input resistance, the $n^{++}$ sheet resistance, and the capacitance of the system. 

\begin{figure}[htb]
\begin{center}
\includegraphics[width=0.95\textwidth]{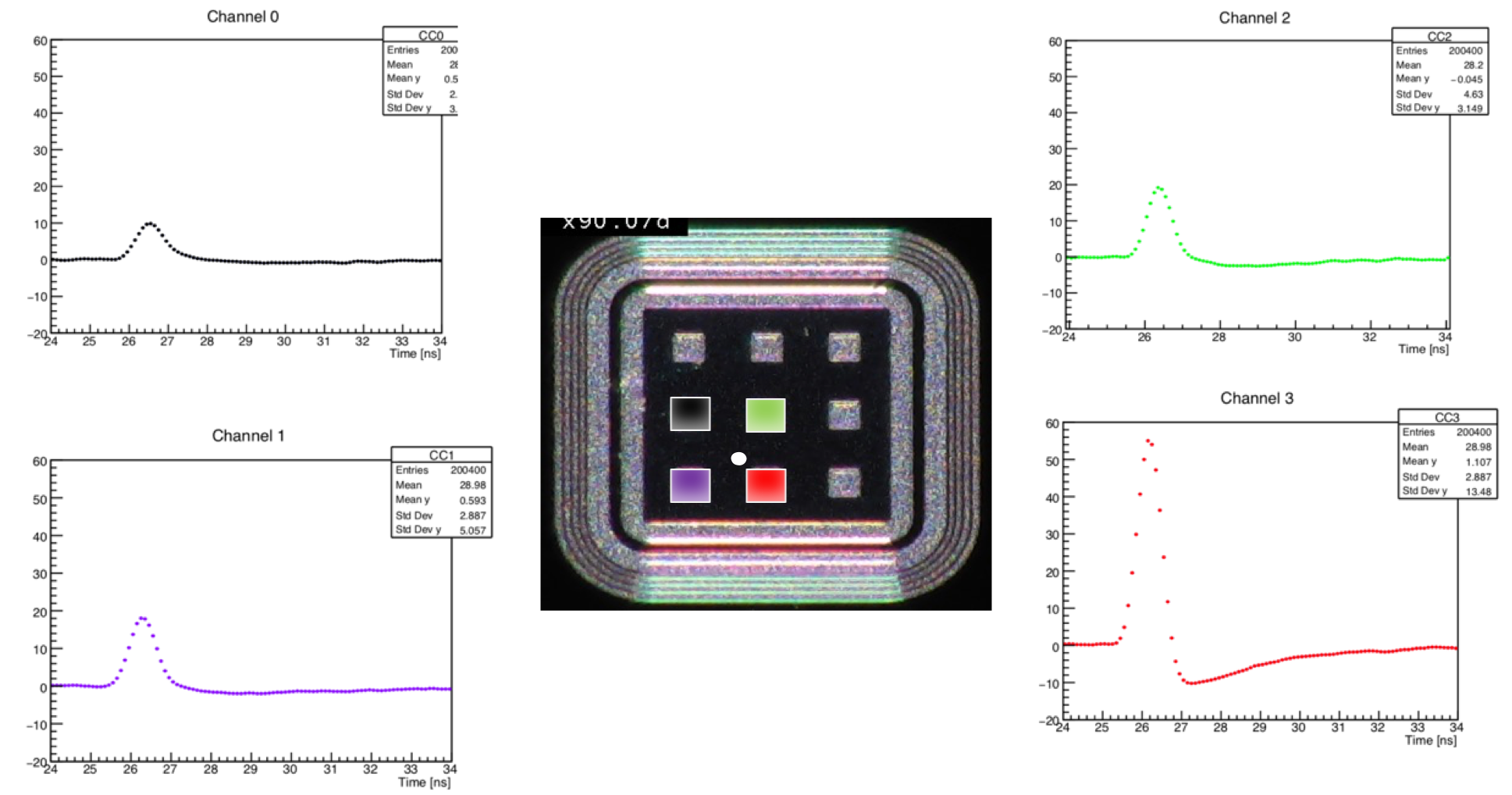}
\caption{Example of signal sharing among 4 pads in AC-LGAD (RSD) sensors: the signal is delayed and attenuated with distance.}
\label{fig:cs}
\end{center}
\end{figure}

The signal is seen on the AC pads with a delay and an attenuation that depends on the distance from the impinging point, as it is reported in Figure~\ref{fig:cs}. The closest pad, marked in red, sees the earliest and largest signal, while the black pad the smallest  and the most delayed one. The  signal is composed by a first lobe, with a shape very similar to that of a standard LGAD, followed by a second lobe, longer and with opposite polarity.  Other important aspects of the signal in AC-LGAD: (i) when summing up all pads, the total amplitude is almost constant regardless of the particle impinging position, (ii) signal attenuation is higher for sensors with a  large fraction of the area covered by metal, attenuation $\propto \; (metal/pitch)^2$,  (iii) the signal delay is about 0.5 - 1.5 ps/$\mu$m, (iv) the signal of particles hitting a metal pad is not shared if the metal pad is larger than 80-100 $\mu m^2$.

In AC-LGAD, the measurement of the arrival time and hit position exploits the  mechanism of charge sharing between multiple pads explained above, reaching  concurrent very good time ($\sigma_t \sim$ 20-30 ps) and position ($\sigma_x \sim 10 \mu$m) resolutions. AC-LGADs are therefore able to achieve a position precision that far exceed that of binary read-out, allowing a strong reduction of the readout channels. As a matter of comparison, a 200 $\mu$m pitch AC-LGAD  has the same spatial resolution of a 25 $\mu$m pitch traditional sensors.  In an hybdrid configuration, this fact has very important consequences as it reduces the number of channels by a factor of $\sim 50$, it allows using more power per channel,  and it provides a lot more real estate per read-out channel. More  results on AC-LGAD (RSD) are presented in~\cite{HSTD12Arci}\cite{HSTD12Pate}.

\section{Conclusions and Outlook}
The characteristics of the silicon tracking detectors  proposed for the  new accelerator facilities are extremely challenging in terms of radiation resistance, spatial and time resolution, power consumption, area, and material budget. A strong R\&D phase is necessary to meet these challenges, together with new ideas in the design of the detectors. Internal gain, introduced in the mainstream silicon design a few years ago with the advent of the LGAD architecture, coupled with the exploitation of the saturation of radiation damage, measured in the last few years,  have the potentiality to help achieving these goals. A new design of silicon detector, the so called AC-LGAD (RSD) architecture, uses charge sharing to achieve the excellent time and spatial resolutions required by the new silicon trackers while reducing the number of channel by more than a factor of 10.  In the next few years the performance of AC-LGAD will be measured and its design optimized. Given their continuos gain layer, the AC-LGAD design is also very promising  for 4D tracking  at small pitch sizes and 100\% fill factor.

\section*{Acknowledgments}

We thank our collaborators within RD50, ATLAS and CMS who participated in the development of UFSD. Part of this work has been financed by the European Union Horizon 2020 Research and Innovation
funding program, under Grant Agreement no.~654168 (AIDA-2020) and Grant Agreement no.~669529 (ERC UFSD669529), and the MIUR via the  Dipartimento di Eccellenza, Physics Dep. of Torino  (ex L. 232/2016, art. 1, cc. 314, 337). The work was supported by the United States Department of Energy, grant DE-FG02-04ER41286.

\section*{References}

\bibliography{NC_bibfile}

\end{document}